\documentclass{PoS}

\def\be{\begin{eqnarray}}
\def\ee{\end{eqnarray}}

\title{Index and overlap construction for staggered fermions}

\ShortTitle{Index and overlap construction for staggered fermions}

\author{\speaker{David H. Adams}\\
Division of Mathematical Sciences,        
Nanyang Technological University, Singapore 637371\\
        E-mail: \email{dhadams@ntu.edu.sg}}

\abstract{Recent developments regarding index and overlap construction
for staggered fermions are reviewed, highlighting the surprising and
unexpected aspects.}

\FullConference{The XXVIII International Symposium on Lattice Field Theory\\
                 June 14-19,2010\\
                 Villasimius, Sardinia Italy}

\begin{document}

\section{Introduction} 

Staggered fermions had long been perceived as disadvantaged compared to 
Wilson fermions regarding the index theorem connection between (would-be) 
zero-modes and gauge field topology. For Wilson fermions, the would-be
zero-modes can be identified as eigenmodes with low-lying real eigenvalues;
these can be assigned chirality $\pm1$ according to the sign of 
$\psi^{\dagger}\gamma_5\psi$, thereby determining an integer-valued index which 
coincides with the topological charge of the background lattice gauge field in
accordance with the index theorem when the gauge field  
is not too rough \cite{SV,Itoh,overlap(H)}. It coincides with the index obtained
from the exact chiral zero-modes of the overlap Dirac operator \cite{Neu}. 
In contrast, for staggered fermions, no way to identify the would-be 
zero-modes was known. They appeared to be mixed in with the other low-lying
modes (all having purely imaginary eigenvalues) \cite{SV,Damgaard-Heller}
and only separating out close to the continuum limit \cite{Davies(index)}. 
It seemed that, away from the continuum limit, the best one could have was
a field-theoretic definition of the staggered fermion index \cite{SV}.
The latter had the disadvantages of being non-integer, requiring a 
renormalization depending on the whole ensemble of lattice gauge fields, and
being significantly less capable than the Wilson fermion index
of maintaining the index theorem in rougher backgrounds 
\cite{SV}.

Recently the consensus viewpoint described above was found to be incorrect:
Staggered fermions do have identifiable would-be zero-modes away from
the continuum limit, with identifiable chiralities and integer-valued
index satisfying the index theorem when the lattice gauge field is not too 
rough \cite{DA(sindex)}. The would-be zero-modes, chiralities and index can 
be identified in a {\em spectral flow} approach based on a new hermitian 
version of the staggered Dirac operator, paralleling the spectral flow 
approach to the index for usual Wilson fermions \cite{Itoh,overlap(H)}.

Further developments along this line have led to a new version of 
overlap fermions built from staggered fermions in place of Wilson fermions
\cite{DA(pairs)}. The construction has the remarkable feature of reducing the
4 fermion flavors described by the staggered fermion to 2 flavors for the
staggered overlap fermion. It turns out that underlying this construction
is a new Wilson-type fermion, obtained by adding a Wilson-type term to the  
staggered fermion, which gives masses $\sim 1/a$ to 2 of the flavors while
leaving the remaining 2 flavors massless. Other Wilson-type terms are also
possible; another one which reduces the flavors from 4 to 1 was subsequently
proposed in \cite{Hoelbling}. Numerical investigations of the 2-flavor
staggered overlap fermion have been reported in \cite{Forcrand(POS)}.
The methods of \cite{DA(sindex),DA(pairs)} were later applied 
to naive fermions and minimally doubled fermions \cite{Creutz1}.

A posteriori, these results and constructions can superficially seem quite 
straightforward. But a priori the odds were very much against any of this 
working out in a sensible way. There were a number of surprises and unexpected
aspects, and these will be highlighted in the present review.

\section{Would-be zero-modes and index of the staggered Dirac operator}

In the continuum setting, the spectral flow perspective on the index 
of the Dirac operator $D$ arises by considering the 
eigenvalues $\{\lambda(m)\}$ of the hermitian operator
\be
H(m)=\gamma_5(D-m)
\label{3}
\ee
The spectral flow is defined as the net 
number of eigenvalues $\lambda(m)$ of $H(m)$ that cross the origin, counted 
with sign $\pm$ depending on the slope of the crossing, as $m$ is varied over 
some range. It can be shown that the spectral flow of $H(m)$  
comes entirely from eigenvalue crossings at $m=0$ and equals minus the 
index of $D$.

In the lattice setting with Wilson fermions,
the spectral flow perspective \cite{Itoh,overlap(H)} is
based on the hermitian lattice analogue of (\ref{3}):
\be
H_W(m)=\gamma_5(D_W-m)
\label{6}
\ee
where $D_W$ is the Wilson Dirac operator. The eigenvalue crossings of 
$H_W(m)$ are in one-to-one correspondence
with {\em real} eigenvalues of $D_W$, and the index of $D_W$ (obtained from
the would-be zero-modes, i.e. the eigenmodes with low-lying real 
eigenvalues) coincides with minus the spectral flow of the low-lying
eigenvalue crossings of $H_W(m)$.
Numerical results illustrating this can be found, e.g., in \cite{Itoh}.
An illustration in the $d\!=\!2$ case is given in Fig.1 where the
eigenvalues of $H_W(m)$ are plotted as functions of $m$.

In the case of staggered fermions, the staggered Dirac operator $D_{st}$ 
is anti-hermitian and therefore all its eigenvalues are purely imaginary.
Hence the identification of would-be zero modes and index in the Wilson
case does not carry over to the staggered case: there are no real eigenvalues,
and in fact the staggered analogue of (\ref{6}), $\Gamma_5(D_{st}-m)$, 
is not even hermitian. The lack of any obvious way to distinguish the would-be 
zero-modes from the other low-lying eigenmodes of $D_{st}$ gave rise to the
consensus viewpoint that staggered fermions are disadvantaged in this 
regard relative to Wilson fermions. 

However, it turns out that there is an alternative spectral flow approach
in the staggered case \cite{DA(sindex)}. Note that in the continuum setting, 
instead of (\ref{3}) one can just as well use the hermitian operator 
$H(m)=iD-m\gamma_5$ for the spectral flow perspective on the index. 
But now the staggered analogue,
\be
H_{st}(m)=iD_{st}-m\Gamma_5
\label{7}
\ee
is also hermitian and so its spectral flow can be considered as well. 
Here $\Gamma_5$ is the analogue of $\gamma_5$ in the 
staggered formulation; it is hermitian and corresponds up to $O(a^2)$ 
discretization errors to $\gamma_5\otimes{\bf 1}$ in the spin$\otimes$flavor 
interpretation \cite{GS}.
Since $H_{st}(0)=iD_{st}$, the would-be zero-modes of $D_{st}$ are able to be 
identified as the eigenmodes with 
eigenvalues $-i\lambda=-i\lambda(0)$ for which the associated flow $\lambda(m)$ 
crosses zero at a low-lying value of $m$. Furthermore, the sign of the slope 
of the crossing is minus the chirality of the would-be zero-mode, and hence 
the index is minus the spectral flow of $H_{st}(m)$ coming from the crossings
at low-lying values of $m$. See \cite{DA(sindex)} for the details of this
identification. 

This way of identifying the would-be zero-modes of $D_{st}$ from the 
low-lying eigenvalue crossings of $H_{st}(m)$ relies on an implicit assumption,
namely that there is a clear separation between the low-lying and high-lying 
crossings. Actually, there is no a priori reason to believe that this 
assumption is true, even in smooth gauge field backgrounds or in the free
field case. In fact one would expect that it is not true. The clear 
separation between low-lying and high-lying crossings in the Wilson case
(as seen in Fig.1) relies crucially on the property $\gamma_5^2={\bf 1}$.
But the staggered version $\Gamma_5$ does not have this property. 
The eigenvalues of $\Gamma_5$ are not $\pm1$ but are distributed throughout
the interval $[-1,1]$. E.g. $0$ is an eigenvalue of $\Gamma_5$ in
the free field case; this can be seen from the free field momentum 
representation of $\Gamma_5^2$ which is $\prod_{\nu}\cos^2(p_{\nu})$.
In light of this one would expect that, even in the 
free field case, the eigenvalue crossings of $H_{st}(m)$ will be an arbitrary 
mess with no clear separation into low-lying and high-lying crossings.

The first and biggest surprise in all this -- a miracle in fact -- is that,
contrary to expectations, the spectral flow of $H_{st}(m)$ does have a clear
separation between low-lying and high-lying eigenvalue crossings, at least
when the gauge field is not too rough. In fact in the free field case there 
are no high-lying crossings at all, cf. the bound (\ref{bound}) below. 
Fig.2 shows the spectral flow in a moderately roughened U(1) background 
with topological charge $Q=1$ on a 2-dimensional lattice. Now
there are high-lying crossings, but they are clearly separated from the 
low-lying ones, so the would-be zero-modes of $D_{st}$, their chiralities,
and index, can be unambiguously identified.

The absence of high-lying eigenvalue crossings for $H_{st}(m)$ in the free
field case can be seen analytically as follows. A simple calculation of 
$H_{st}(m)^2$ in the free field momentum representation gives
\be
\hat{H}_{st}(m)^2=\sum_{\mu}\sin^2(p_{\mu})+m^2\prod_{\nu}\cos^2(p_{\nu})
\label{8}
\ee
Set $s_{\mu}=\sin(p_{\mu})$, $c_{\nu}=\cos(p_{\nu})$. Then, in the
case of 2 spacetime dimensions, starting from 
$\hat{H}_{st}(m)^2=s_1^2+s_2^2+m^2(1-s_1^2)(1-s_2^2)$, we find
\be
\hat{H}_{st}(m)^2=m^2+(1-m^2)(s_1^2+s_2^2)+m^2s_1^2s_2^2
\quad \ge\;m^2\quad \mbox{for $\ 0\le|m|\le1$}
\label{9}
\ee
and
\be
\hat{H}_{st}(m)^2=1+s_1^2s_2^2+(m^2-1)(1-s_1^2)(1-s_2^2)
\quad \ge\;1\quad \mbox{for $\ |m|\ge1$}
\label{10}
\ee
Note that both of these bounds are saturated. Identical bounds can be derived
in the $d=4$ case, although the derivations are more complicated. Hence
in the free field case, for both $d=2$ and $d=4$ dimensions (and probably 
also for higher dimensions), we have
\be
H_{st}(m)_{free}^2\;\ge\;
\left\{
\begin{array}{ll}
m^2 & \quad \mbox{for $|m|\le1$}\\
1 & \quad \mbox{for $|m|\ge1$}\\
\end{array} \right.
\label{bound}
\ee
This bound has a generalization to the case of gauge fields satisfying 
an ``admissibility condition'' on the plaquettes that implies a separation 
between low-lying
and high-lying eigenvalue crossings for $H_{st}(m)$ when the $\epsilon$
in the condition is sufficiently small -- see \cite{DA(sindex)} for
details. Thus the situation is analogous to the Wilson case where the 
admissibility condition guarantees the separation between low-lying and
high-lying crossings for the hermitian Wilson operator $H_W(m)$
\cite{L-Neu(bound)}.

Comparing Fig.'s 1 and 2 we see that the form 
of the spectral flow of $H_{st}(m)$ is very different from
the Wilson case, and the separation between the 
low-lying and high-lying crossings is much larger. However, this is not one 
of the surprises alluded to in the abstract. Instead, the surprise here is 
that there is no surprise -- the staggered spectral flow has the same form as 
in the Wilson case once the correct interpretation of the hermitian staggered 
operator $H_{st}(m)$ is identified. It turns out that the parameter
$m$ in the staggered case should be identified not with the corresponding
$m$ in $H_W(m)$ but with the Wilson parameter $r$ in the Wilson case. This
will be explained further below; see Fig.3.   

Since the staggered fermion in $d$ dimensions describes $2^{d/2}$ flavors,
the index theorem in this case should be 
$\mbox{index}(D_{st})=2^{d/2}(-1)^{d/2}Q$. This is confirmed by 
numerical results in smooth enough backgrounds. E.g. in Fig.2
the two positive slope low-lying
crossings in the $Q=1$ background in 2 dimensions imply that
the index is $-2$ in accordance with the index theorem.
The eigenvalues of $D_{st}$ in this background correspond to $m=0$ in Fig.2.
The would-be zero-modes of $D_{st}$ can be identified as the two eigenmodes 
with the low-lying eigenvalues belonging to the two eigenvalue flows that 
cross the origin.  

\begin{figure}[ht]
\begin{minipage}[b]{0.4\linewidth}
\centering
\includegraphics[width=2.5in,clip]{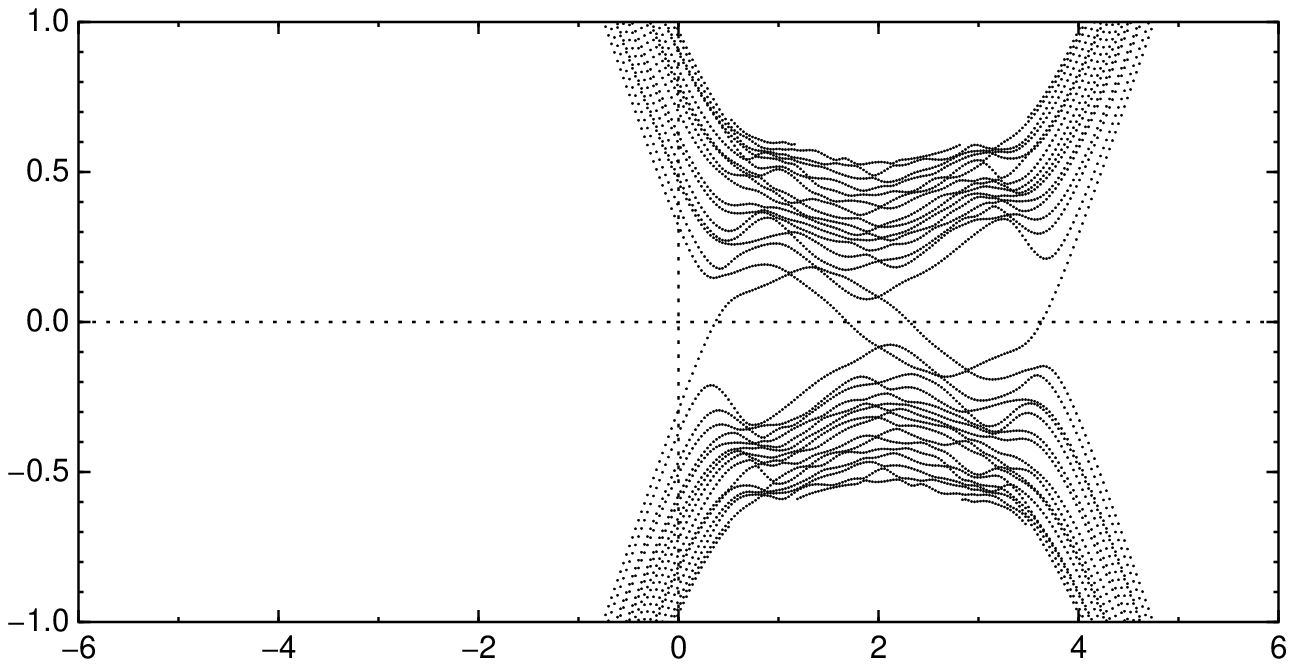}
\caption{\label{pos1eps} Eigenvalue flow of $H_W(m)$ 
in a $Q\!=\!1$ background on a 2-dim lattice.}
\end{minipage}
\hspace{1.5cm}
\begin{minipage}[b]{0.4\linewidth}
\centering
\includegraphics[width=2.5in,clip]{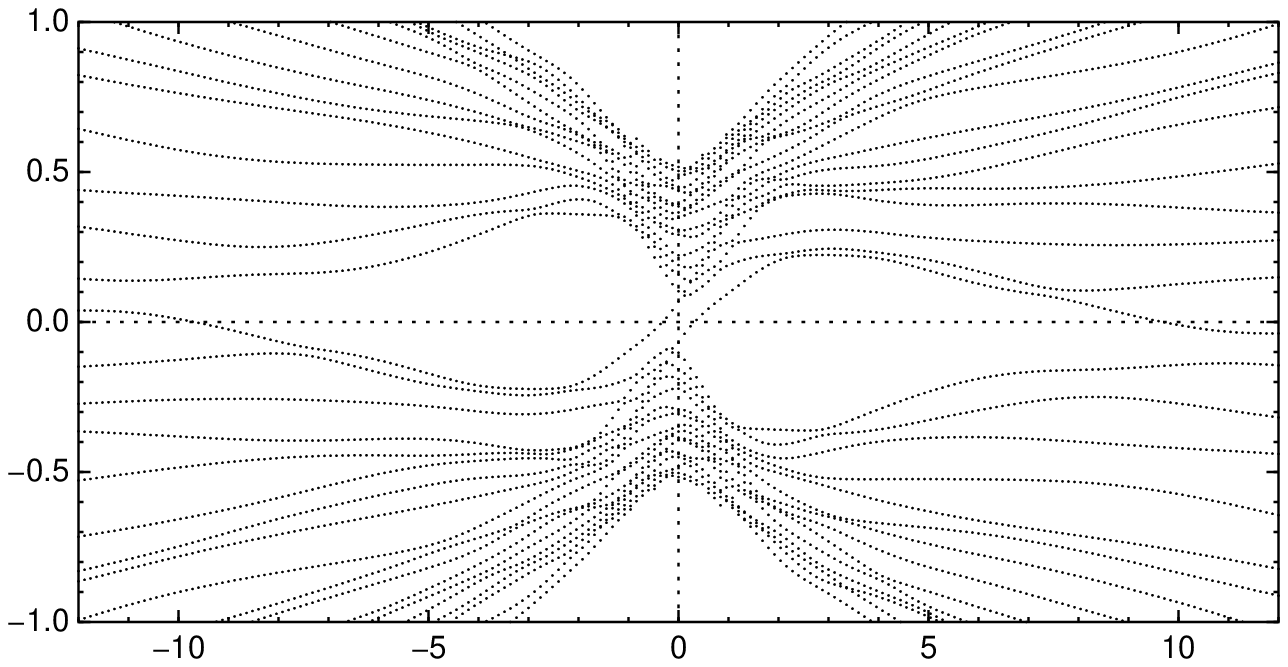}
\caption{\label{pos2eps} Eigenvalue flow of $H_{st}(m)$ in the same 
$Q\!=\!1$ background as Fig. 1.}
\end{minipage}
\end{figure} 

\section{Staggered overlap construction}

In the Wilson case, the spectral flow perspective on the index leads to
\be
\mbox{index}(D_W)=
{\textstyle -\frac{1}{2}}\mbox{Tr}\frac{H_W(m_0)}{\sqrt{H_W(m_0)^2}}
=\mbox{index}(D_{ov})
\label{10a}
\ee
for any $m_0$ in the region between where the low-lying and high-lying 
eigenvalue crossings of $H_W(m)$ occur (e.g. $m_0=1/a$) 
\cite{overlap(H),Neu}.
Here $D_{ov}=\frac{1}{a}\Big(1+\gamma_5\frac{H_W(m_0)}{\sqrt{H_W(m_0)^2}}\Big)$
is the overlap Dirac operator \cite{Neu}. 
The intimate connection between the Wilson index, the hermitian
operator $H_W(m)$ and the overlap Dirac operator suggests there may
exist a staggered version of the overlap Dirac operator connected to the 
staggered index and staggered hermitian operator $H_{st}(m)$ discussed above.
But there is a problem: The connections and properties of the overlap
Dirac operator in the Wilson case rely crucially on the property
$\gamma_5^2={\bf 1}$. (E.g. without it the GW relation would not hold and
$D_{ov}$  would not have exact zero-modes.) Therefore the natural replacement
$\gamma_5\to\Gamma_5$ is not possible when constructing the overlap operator in
the staggered case since $\Gamma_5^2\ne{\bf 1}$. 
Attempts to construct different versions of $\gamma_5$ in the staggered case
which do satisfy $\gamma_5^2={\bf 1}$ invariably lead to unnatural, problematic 
operators which violate either lattice rotation invariance or gauge invariance
\cite{DA(unpublished)}. 

This ``$\Gamma_5^2\ne{\bf 1}$ problem'' initially appears insurmountable,
but the second main surprise in all this is that it
does have a solution \cite{DA(pairs)}. The theoretical idea behind 
the solution is as follows. In the staggered setting there is a naturally 
arising operator which squares to the identity, namely $\Gamma_{55}$,
acting on the staggered fermion fields by 
$\Gamma_{55}\chi(x)=(-1)^{n_1+\dots+n_d}\chi(x)$. It has the spin$\otimes$flavor
interpretation $\gamma_5\otimes\gamma_5$, which is not what we want. But if
the staggered overlap construction can be set up such that the physical 
flavors are those with positive flavor-chirality under 
${\bf 1}\otimes\gamma_5$ then $\gamma_5\otimes\gamma_5$ will be the same as 
$\gamma_5\otimes{\bf 1}$ on the physical flavors, and then $\Gamma_{55}$ 
may be used for the role of $\gamma_5$ in the staggered overlap construction.
    
In fact this can be achieved simply by replacing $\gamma_5\to\Gamma_{55}$
and $H_W \to H_{st}$ in the overlap formula for $D_{ov}$. The key observation
is that in this construction we have
$\Gamma_{55}H_{st}(m_0)=i\Gamma_{55}D_{st}-m_0\Gamma_{55}\Gamma_5$ and 
$\Gamma_{55}\Gamma_5$ has the spin$\otimes$flavor interpretation 
$(\gamma_5\otimes\gamma_5)(\gamma_5\otimes{\bf 1})
={\bf 1}\otimes\gamma_5$ up to $O(a^2)$ effects. 
Thus the 2 fermion flavors with positive
flavor-chirality get a negative mass from the 
$-m_0\Gamma_{55}\Gamma_5$ term, and hence become massless flavors of the
staggered overlap fermion, while the 2 flavors with negative 
flavor-chirality get positive masses from this term and hence become 
heavy, decoupling flavors of the staggered overlap fermion with masses 
$\sim 1/a$, just like the ''doubler'' species in the usual overlap fermion
construction. 
Note that the exact {\em flavored} chiral symmetry $\{D_{st}\,,\Gamma_{55}\}=0$
of the staggered fermion hereby becomes an exact {\em unflavored} GW chiral 
symmetry $\{D_{sov}\,,\Gamma_{55}\}=aD_{sov}\Gamma_{55}D_{sov}$ of the resulting
staggered overlap Dirac operator $D_{sov}$.
Moreover, a staggered version of the index relations (\ref{10a}) holds 
\cite{DA(sindex),DA(pairs)}: 
$\frac{1}{2}\mbox{index}(D_{st})=
-\frac{1}{2}\mbox{Tr}\frac{H_{st}(m_0)}{\sqrt{H_{st}(m_0)^2}}
=\mbox{index}(D_{sov})$.
The factor $\frac{1}{2}$ multiplying index$(D_{st})$ reflects the reduction 
from 4 to 2 flavors in the staggered overlap fermion.

The interpretation of the staggered overlap fermion becomes more
straightforward if we change the hermitian staggered operator by
$H_{st}(m)\;\to\;\Gamma_{55}D_{st}-m\Gamma_5
=\Gamma_{55}(D_{st}-m\Gamma_{55}\Gamma_5)$.
As mentioned in \cite{DA(pairs)}, this operator is closely related to, and 
has the same eigenvalue spectrum as, the previous operator $iD_{st}-m\Gamma_5$.
Everything in the preceding continues to hold with this new $H_{st}(m)$.
The staggered overlap Dirac operator takes a more recognizable form 
though: it can now be written as 
\be
D_{sov}=\frac{1}{a}\Big(1+(D_{st}-m_0\Gamma_{55}\Gamma_5)
\frac{1}{\sqrt{(D_{st}-m_0\Gamma_{55}\Gamma_5)^{\dagger}
(D_{st}-m_0\Gamma_{55}\Gamma_5)}}\Big)
\label{10e}
\ee
From this we see that underlying the staggered overlap construction is
a new {\em staggered version of Wilson fermions} with the Dirac operator
\be
D_{sW}=D_{st}+W_{st}\quad,\quad W_{st}=\frac{r}{a}(1-\Gamma_{55}\Gamma_5).
\label{10f}
\ee
The ``Wilson term'' $W_{st}$ decouples the negative
flavor-chirality modes by giving them mass $2r/a$ while keeping the
two positive flavor-chirality modes as the physical modes.
Hence $D_{sW}$ describes two physical quark flavors on which 
$\Gamma_5=\Gamma_{55}$ up to $O(a)$ effects. It has the $\Gamma_{55}$ 
hermiticity $D_{sW}^{\dagger}=\Gamma_{55}D_{sW}\Gamma_{55}$.
A 2-flavor overlap fermion can then be obtained by taking $D_{sW}-m$ with 
$m=\frac{r\rho}{a}\,$, $\rho\in(0,2)$ as the kernel in the usual overlap 
construction. For $\frac{r}{a}=m_0$ and $\rho=1$ this is precisely the 
2-flavor staggered overlap Dirac operator $D_{sov}$ obtained above in
(\ref{10e}). But now we see that it can be
generalized to any $\rho\in(0,2)$. Furthermore, the role of the parameter
$m_0$ in the staggered overlap construction is hereby clarified: it is 
analogous to the Wilson parameter in the usual overlap construction.

The general staggered overlap operator can also be expressed as
$D_{sov}=\frac{1}{a}(1+\Gamma_{55}\frac{H_{sW}(\rho)}{\sqrt{H_{sW}(\rho)^2}})$
where $H_{sW}$ is another hermitian staggered operator given by\footnote{We 
set $r=1$ and use lattice units to get the second equality.}
\be
H_{sW}(m)=\Gamma_{55}(D_{sW}-m)=\Gamma_{55}D_{st}-\Gamma_5+(1-m)\Gamma_{55}
\label{10g}
\ee
This is the true analogue of the hermitian Wilson operator 
$H_W(m)=\gamma_5(D_W-m)$, and its spectral flow has a similar 
form (Fig.4).\footnote{In 2 dimensions the staggered Wilson fermion has one
physical flavor and one doubler whereas the usual Wilson fermion has three
doublers. This explains why there is one high-lying eigenvalue
crossing in Fig.4 and three high-lying crossings in Fig.1.} 
On the other hand, the spectral flow of the hermitian Wilson
operator as a function of the Wilson parameter $r$, with fixed $m=1$
(Fig.3), has a
similar form to the spectral flow of our previous hermitian staggered
operator $H_{st}(m)$ as a function of $m$ (Fig.2) 
as anticipated.\footnote{Note from (\ref{10g}) that 
the flow parameter $m$ of $H_{sW}(m)$ is 
multiplied onto $\Gamma_{55}$, whereas for $H_{st}(m)$ it is multiplied
onto $\Gamma_5$. Thus the gauge field topology can be probed by a hermitian
operator whose varying term is the flavor non-singlet $m\Gamma_{55}$, 
contrary to an incorrect claim in \cite{Forcrand(POS)}.} 

\begin{figure}[ht]
\begin{minipage}[b]{0.4\linewidth}
\centering
\includegraphics[width=2.5in,clip]{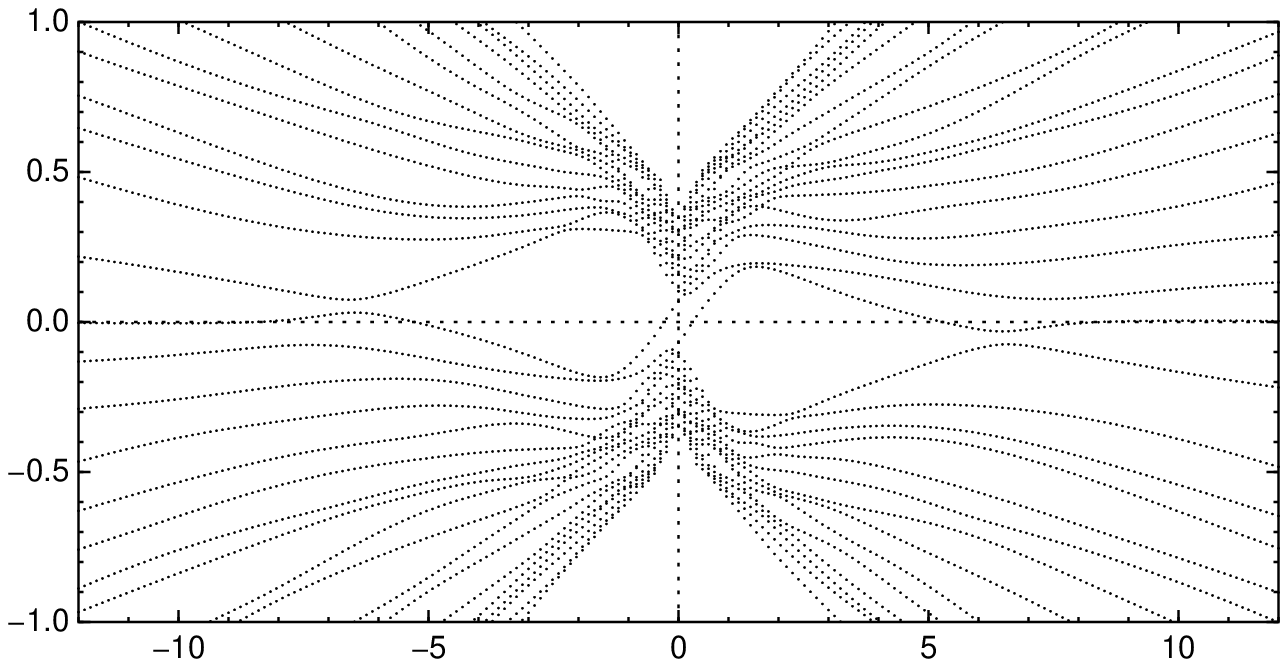}
\caption{\label{pos3eps} Eigenvalue flow of $H_W(m=1)$ 
as a function of the Wilson parameter $r$.}
\end{minipage}
\hspace{1.5cm}
\begin{minipage}[b]{0.4\linewidth}
\centering
\includegraphics[width=2.5in,clip]{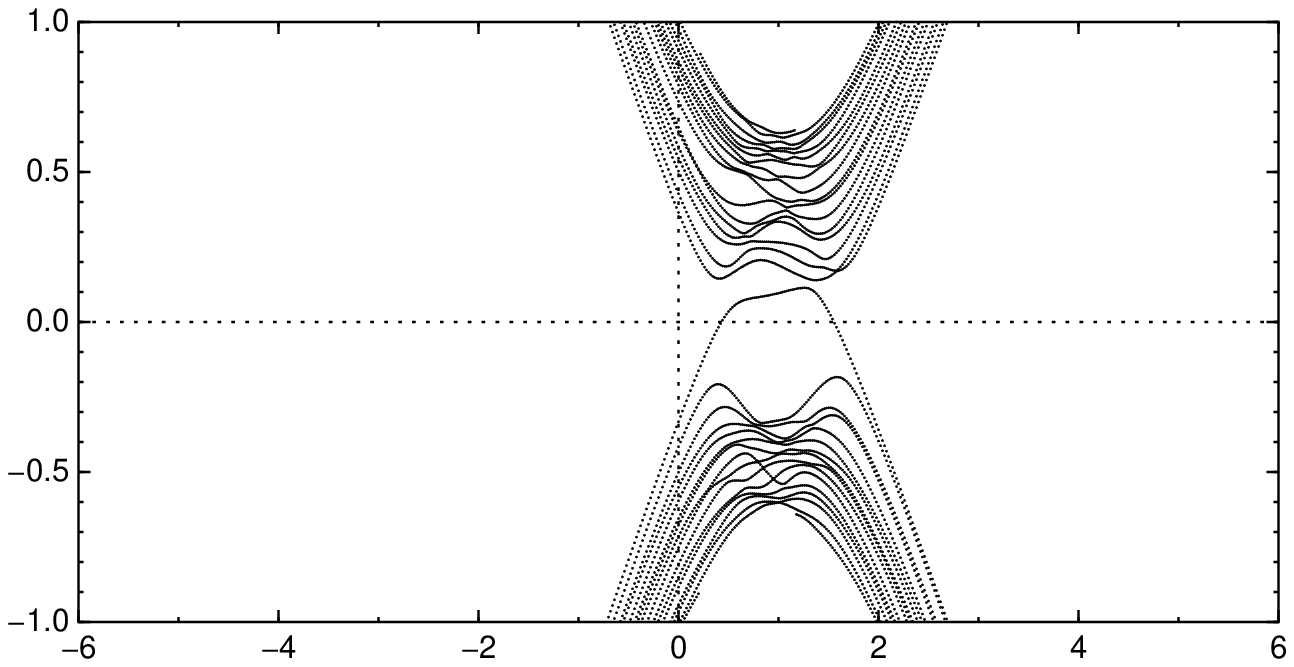}
\caption{\label{pos4eps} Eigenvalue flow of $H_{sW}(m)$ in the same 
$Q\!=\!1$ background as Fig.s 1,2,3.}
\end{minipage}
\end{figure}

To summarize, three new lattice fermion formulations have been introduced:
staggered versions of Wilson fermions, domain wall 
fermions\footnote{Staggered domain wall fermions were not discussed here; 
see \cite{DA(pairs)}.} and overlap fermions. 
The relation of ordinary staggered fermions to these
staggered-based formulations is analogous to the relation of naive fermions
to the Wilson-based formulations. The crucial property of the
``Wilson term'' in the staggered version of Wilson fermions is that it
splits the 4 flavors into pairs with positive and negative flavor-chirality,
giving a mass $\sim 1/a$ to the latter while leaving
the former pair as the massless physical fermions. 
This allows $\Gamma_{55}\sim\gamma_5\otimes\gamma_5$ to be used in place of 
$\Gamma_5$, thus overcoming the ``$\Gamma_5^2\ne{\bf 1}$ problem''. 
The staggered versions of domain wall and overlap fermions can then be
constructed simply by replacing $\gamma_5\to\Gamma_{55}$ and $D_W\to D_{sW}$
in the usual formulations.

Each of the staggered-based formulations is a new alternative and competitor
to the corresponding Wilson-based formulation, and should be more
computationally efficient since the constructions start from staggered rather
than naive fermions. However, the gain in efficiency is reduced because
the staggered ``Wilson term'' is less local; it involves $\Gamma_5$ which
is a 4-link operator \cite{Forcrand(POS)}. This reduction in efficiency may
possibly be ameliorated by smearing the links \cite{Hoelbling}.

Much remains to be done to clarify the theoretical and practical 
properties of the new staggered-based formulations. They break some of the
symmetries of the original staggered fermion and the consequences of this
need to be investigated and clarified. This may be done in lattice 
perturbation theory and also non-perturbatively by numerical lattice QCD
calculations. E.g. it should be checked for the staggered Wilson fermion  
that a chiral (massless) limit can be reached by tuning the bare quark 
mass.\footnote{A step in this direction was made in \cite{Creutz2}.}
For the staggered domain wall fermion and staggered overlap fermion it
should be checked that the lattice QCD theory has an approximately 
and exactly massless phase, respectively. Work on this is currently underway.

\end{document}